\documentclass[prm,superscriptaddress,twocolumn]{revtex4-2}
\usepackage{color}
\usepackage{chemformula} 
\usepackage{placeins}

\usepackage{graphicx}
\usepackage{amsmath}
\usepackage{siunitx}
\usepackage[T1]{fontenc}

\usepackage{amsfonts} 

\usepackage{hyperref}  

\begin{document}

\title{Quantum confinement in semiconductor random alloys: a case study on Si/SiGe/Si}

\newcommand{\ZfM}{Chemnitz University of Technology, Center for Microtechnologies, 09126 Chemnitz, Germany\\}
\newcommand{\ENAS}{Fraunhofer Institute for Electronic Nano Systems, 09126 Chemnitz, Germany\\}
\newcommand{\MAIN}{Chemnitz University of Technology, Center for Materials, Architectures and Integration of Nanomembranes, 09126 Chemnitz, Germany\\}
\newcommand{\IoP}{Chemnitz University of Technology, Institute of Physics, 09126 Chemnitz, Germany\\}

\author{Daniel Dick}
\email{daniel.dick@zfm.tu-chemnitz.de}

\author{Florian Fuchs}
\author{Jörg Schuster}
\affiliation{\ZfM}
\affiliation{\ENAS}
\affiliation{\MAIN}

\author{Sibylle Gemming}
\affiliation{\MAIN}
\affiliation{\IoP}


\begin{abstract}
Local composition fluctuations in random alloys become crucial when one or more dimensions are reduced to the nanoscale. Using extended Hückel theory, we study the semiconductor random alloy SiGe sandwiched between Si due to its relevance for transistor devices. We evaluate the effects of the alloy composition, layer thickness, and local  fluctuations of the Ge concentration on the band alignment and the band gap. The results are compared with the finite quantum well model. That model captures the essential physics and can act as a computationally faster alternative. 

\end{abstract}

\date{\today}
\maketitle

\section{Introduction}
Semiconductor alloys allow tuning electronic properties, such as the band gap, by changing the alloy composition. This adjustability makes them attractive materials for a wide range of applications. Thin layers of SiGe, for example, are of interest in semiconductor devices such as heterojunction bipolar transistors \cite{Rinaldi2022, Phillips_2021, Schroter_2011}, field-effect transistors \cite{Takagi_2011, Chu_2023, Fuchsberger_2023,Khatami_2009} and quantum cascade lasers \cite{Xia_2011}. 

With the ongoing downscaling of such devices, quantum confinement increases the band gap. Additionally, local fluctuations in alloy composition have an increasing impact as the layer thickness is reduced. 
Studying that effect therefore requires a method by which fluctuations on the atomic scale can be considered.  

Quantum confinement in Si-based nanowires and nanoslabs has been studied previously \cite{Fuchs_2019,Joseph_2023,Tan_2015,Kharche_2008}, but such nanostructures are often simulated in vacuum, without the focus on layer stacks, and the effect of fluctuations in the atomic structure has not been intensively investigated yet \cite{Iori_2014,Martinez_Blanque_2014}.

In the present study, we address the topics of quantum confinement in ultra-thin SiGe layers in-between Si and the role of local stoichiometry fluctuations by simulating different atomic structures and layer thickness values using extended Hückel theory (EHT).

The paper is structured as follows: 
Following this overview, Section~\ref{sec:methodology} introduces the methodology of this study and the simulated atomic structure models.
Afterwards, in Section~\ref{sec:results}, we showcase the calculated band gaps of nm-thin SiGe layers, sandwiched between Si, in dependence on the layer thickness and the alloy composition. 
We consider the topic of band alignment and give a parameterized description of the conduction band and valence band edges with regard to the Ge concentration.
Next, we study the effect of quantum confinement and compare EHT results with the models of the finite and the infinite potential well. 
Finally, we quantify the effect of local fluctuations of the Ge content.

\section{Methodology}
\label{sec:methodology}
In view of the large number of atoms to be considered in few-nm-thick layers, the computationally efficient extended Hückel theory (EHT) \cite{Hoffmann1963,Cerda2000,Stokbro_2010} was chosen for the present study. We employ a parameterization of Si and Ge that was shown to be suited for describing the properties of SiGe alloys by comparison with self-consistent density functional theory \cite{Dick2025}. 
The vacuum level of Ge was shifted by $-0.55$ eV compared to the cited parameterization to more accurately describe the band offset (see section~\ref{sec:alignment} for further discussion).
The calculations are done using Synopsys QuantumATK X-2025.06 \cite{Smidstrup2019,Stokbro_2010}.

\begin{figure}[hbt]
	\centering
	\includegraphics[width=.4\textwidth]{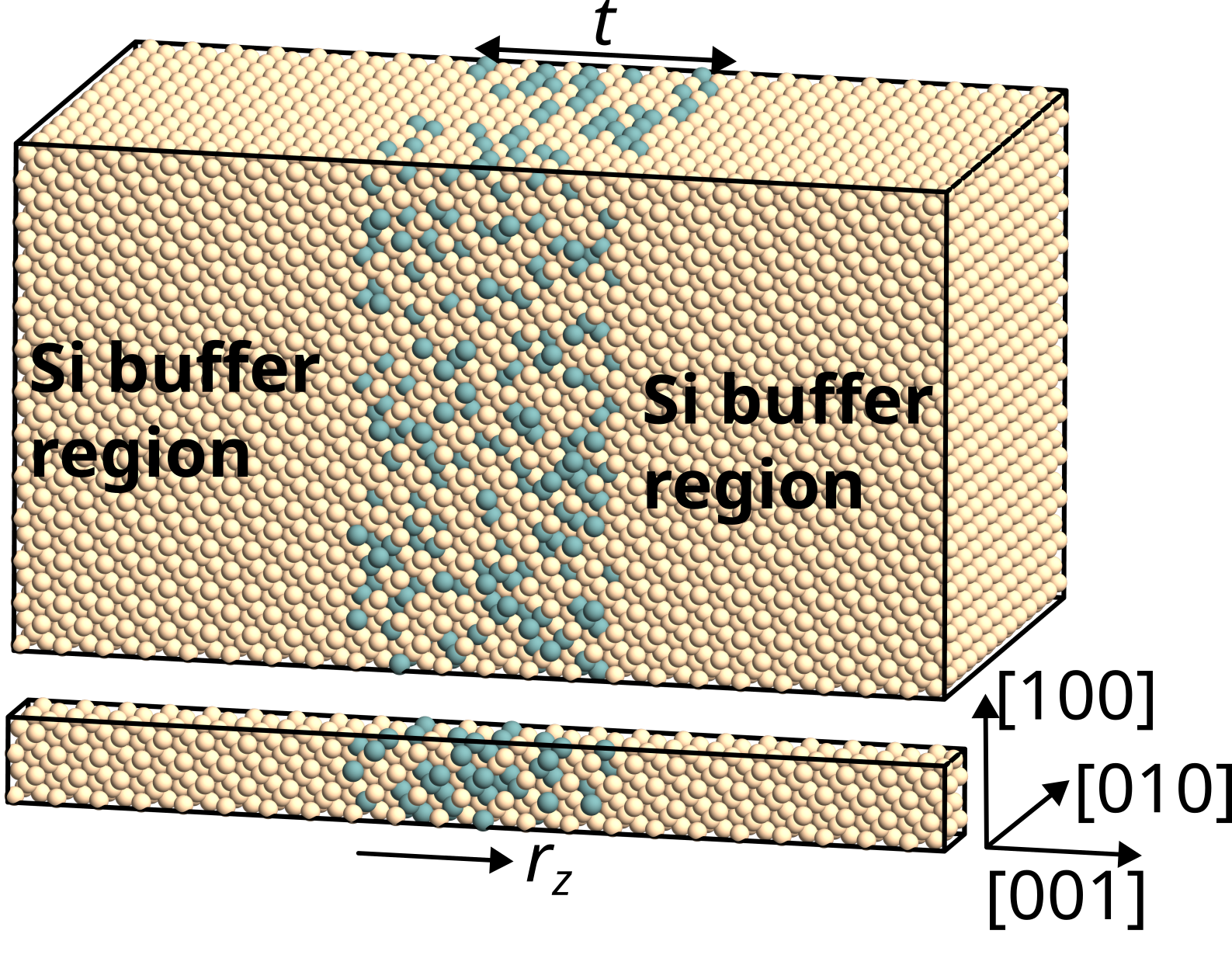}
	\caption{The supercell of a 3~nm SiGe layer used for the structural relaxation, with randomly distributed Si and Ge atoms and an average Ge concentration of 30\% (top). It is divided into 36 sub-cells (bottom) of appropriate size for EHT calculations.}
	\label{fig:supercell}
\end{figure}

The simulation cell is assembled in the following way: First, a single layer is defined by a $12\times12\times1$ grid of conventional unit cells (roughly 6.5~nm $\times$ 6.5~nm $\times$ 0.5~nm). Each atom is randomly selected as either Si or Ge, depending on the chosen alloy composition. 
Multiples of such layers are then stacked on top of each other in $r_z$-direction until the chosen layer thickness $t$ is reached. 
Then, this cell is sandwiched between two Si layers of the same cross-sectional area and a thickness of 8 conventional unit cells (roughly 6.5~nm $\times$ 6.5~nm $\times$ 4.3 nm) each, acting as a separator between periodic images of the SiGe layers in the $r_z$-direction (see Figure~\ref{fig:supercell}). 
For a comparison to bulk alloys, additional simulations without the Si layers are performed as well.

The supercell with more than 20,000 atoms is relaxed using a parameterization of the Stillinger-Weber potential \cite{Stillinger1985} by Laradji et al.~\cite{Laradji1995}. The lattice constants in the in-plane directions, $r_x$ and $r_y$, are fixed to the lattice constant of pristine Si, resembling a SiGe layer inside a pure Si matrix. Only the lattice constant in the confined out-of-plane direction changes during the relaxation step. The lattice constant, which is not further discussed in this study, closely follows Vegard's law, in agreement with recent DFT simulations \cite{Fuchs_2025,Tunica_2026}.

The relaxed supercell is then divided into 36 cells of $\sim$~500~-~1000 atoms (depending on layer width, see Figure~\ref{fig:supercell}). 
Because of the random distribution of Si and Ge atoms in the supercell, each of those sub-cells has a slightly different alloy composition. This allows to quantify the effect of local stoichiometry fluctuations. 
There is a small difference between the mean Ge concentration of all cells $\langle x \rangle$ and the chosen target Ge content $x$, most notably for the thinnest layers studied in this work. This leads to an uncertainty of $\sim$~2~meV of the mean band gap, which is within the accuracy of our method. Appendix~\ref{sec:error_estimation} describes this in more detail.

\begin{figure}[hbt]
	\centering
	\includegraphics{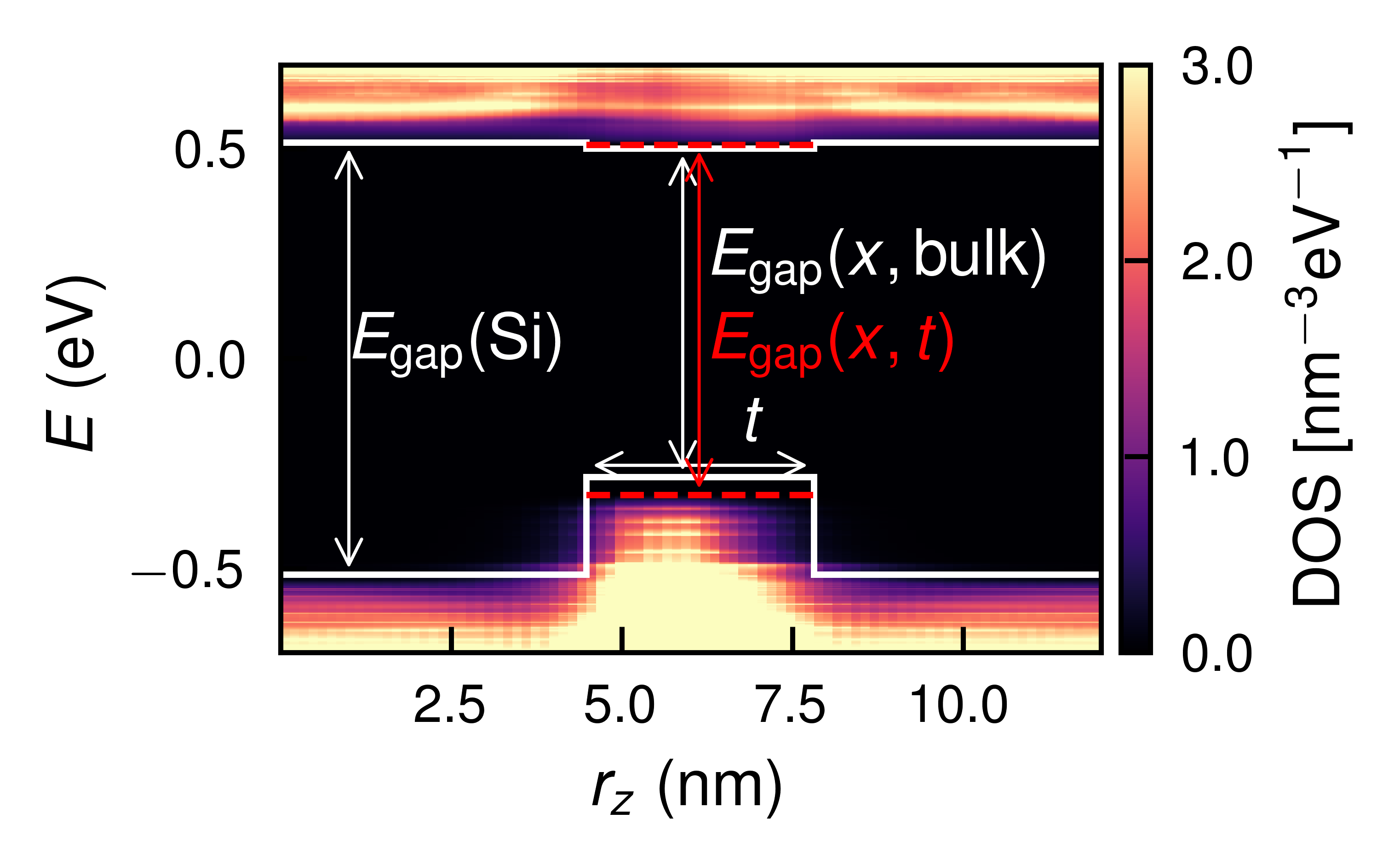}
	\caption{The band alignment of the structure depicted on the bottom of Figure~\ref{fig:supercell} and its atom-projected density of states (dos). The band gap of the layer, $E_{\text{gap}}(x,t)$, is larger than the corresponding band gap of the bulk material, $E_{\text{gap}}(x,\text{bulk})$, due to quantum confinement.}
	\label{fig:LDOS}
\end{figure}

As indicated in Figure~\ref{fig:LDOS}, the band gap of each structure is calculated as the smallest difference between valence band and conduction band edge. As shown by the local density of states, the band alignment between Si and SiGe leads to a quantum well for both electrons and holes.

\section{Results and discussion}
\label{sec:results}

\begin{figure*}[hbt]
	\centering
	\includegraphics{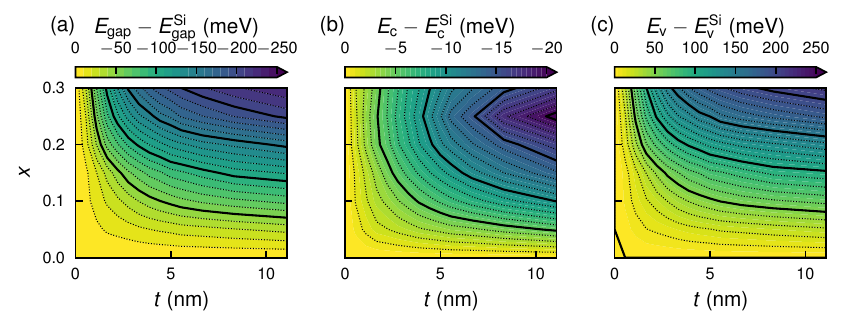}
	\caption{Band gap $E_{\text{gap}}$ (a), conduction band edge $E_{\text{c}}$ (b), and valence band edge $E_{\text{v}}$ (c) of SiGe layers of varying thickness $t$ and Ge content $x$. The colorbar scale in (b) was adjusted to better visualize the change in conduction band energy, which is one order of magnitude smaller than the change in the valence band.}
	\label{fig:band_gap_colormap}
\end{figure*}

Figure~\ref{fig:band_gap_colormap}a) shows how the band gap of SiGe alloys changes with the Ge content and the layer thickness. We focus on the technologically relevant alloy compositions of up to 30\%~Ge and average the band gap values of 36 atomic structures each. 
Two trends can be observed. First, the band gap decreases with higher Ge content. This finding agrees well with prior research on bulk SiGe alloys on Si substrate \cite{Fuchs_2025, Dick2025, King1989, Lang1985, Braunstein1958, Weber_1989}. Secondly, the band gap increases with decreasing layer thickness due to quantum confinement. This counteracts to the effect of Ge content and needs to be considered in smaller devices. For example, as indicated by the contour lines in the figure, a 3~nm thin layer with 30\% Ge has approximately the same band gap as a bulk alloy with a Ge content of 20\%. 

Figures~\ref{fig:band_gap_colormap}b) and~\ref{fig:band_gap_colormap}c) showcase the position of the conduction band edge ($E_\text{c}$) and valence band edge ($E_\text{v}$). The conduction band offset is significantly smaller, leading to a weak confinement of the electrons. The change in band gap is therefore primarily dictated by the confinement of holes in the present model.

\subsection{Band alignment}
\label{sec:alignment}

Both type-I and type-II band alignment of SiGe on Si have been discussed in the literature \cite{Shiraki_2005, Houghton_1995,Thewalt_1997,Virgilio_2006}, and are correlated with doping, strain and alloy composition. It is agreed that for weakly doped alloys, the conduction band offset between Si and SiGe is small compared to the valence band offset.
Our simulation assumes a type-I band alignment in the simulated range of Ge content, which proved successful for device simulations \cite{Jungemann2003_chapter6}.

\begin{figure}[hbt]
	\centering
	\includegraphics{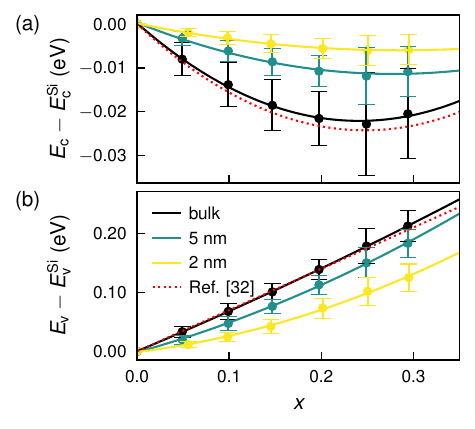}
	
	\caption{ Conduction band (a) and valence band (b) edges as a function of the Ge content $x$ interpolated by a parabolic fit. Results of thin layers and the bulk material are compared to Jungemann and Meinerzhagen \cite{Jungemann2003_chapter6}. We include the standard deviation from our ensemble of 36 different atomic structures. 
	}
	\label{fig:bandedges}
\end{figure}

Figure~\ref{fig:bandedges} illustrates how the position of the band edges changes in the thin layers as a function of the Ge content. The standard deviation of the ensemble of 36 atomic structures, shown in this and following figures, is not to be interpreted as an error of our simulations, but instead refers to the spread in local electronic properties.

In contrast to the reference, we observe a light bowing in the valence band for the bulk material. This bowing increases for thinner layers, while the curve for the conduction band becomes flatter. 

The dependence of the band edges of SiGe alloys on the Ge content $x$ is typically described by a polynomial to the 2nd degree in $x$ in the form
\begin{subequations}
	\label{eq:quadratic_all} 
	\begin{eqnarray}
		E_{\text{c}}(x) = E^{\text{Si}}_{\text{c}} + a x + b x^2 \quad ,
		\label{eq:quadratic_CB}
		\\
		E_{\text{v}}(x) = E^{\text{Si}}_{\text{v}} + c x + d x^2  \quad .
		\label{eq:quadratic_VB}
	\end{eqnarray}
\end{subequations}

The band gap is then obtained as the difference of the conduction and valence band energies.
Table~\ref{tab:band_edges} lists the parameters $a, b, c, d$ of the corresponding interpolations in Figure~\ref{fig:bandedges}. With decreasing film thickness the parameters $a$, $b$, and $c$ decrease in absolute value, while $d$ increases. For the conduction band the ratio $b:a$ remains approximately 2:1, thus the quadratic $x$ dependence dominates here for all film thicknesses, but converges to the limit of non-interacting Ge defect layers for very thin layers.
For the valence band, the composition dependence changes with the layer thickness. In the case of bulk SiGe, the linear term clearly prevails. With decreasing layer thickness, however, the quadratic term dominates similar as for the conduction band. Overall, $c$ and $d$ are larger than $a$ and $b$, indicating that defect levels in the valence band interact more strongly.

\begin{table}[hbt]
	\centering
	\begin{tabular}{c|c|c|c|c}
		thickness &  $a$ (eV) & $b$ (eV) & $c$ (eV) & $d$ (eV)  \\
		\hline
		bulk \cite{Jungemann2003_chapter6} & -0.196 & 0.396 & 0.7 & 0.0 \\
		bulk & -0.185 & 0.383 & 0.658 & 0.224 \\
		5 nm & -0.084 & 0.152 & 0.436 & 0.674 \\
		2 nm & -0.043 & 0.075 & 0.192 & 0.831 \\
	\end{tabular}
	\caption{ Parameters of the 2nd order fit to the conduction band and valence band edges. Results of the bulk material are compared to parameters by Jungemann and Meinerzhagen \cite{Jungemann2003_chapter6}, based on results by Schäffler \cite{Schaeffler_1997}.}
	\label{tab:band_edges}
\end{table}

\subsection{Quantum confinement}
\begin{figure}[hbt]
	\centering
	\includegraphics{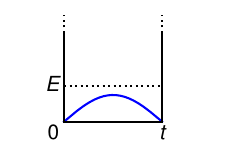}
	\includegraphics{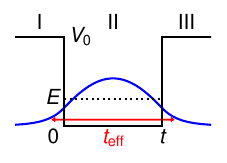}
	\caption{Infinite and finite square well and the ground state wave function. The wave function penetrates the walls of the finite well with exponential decay, giving rise to the concept of an effective layer thickness $t_{\text{eff}}>t$.}
	\label{fig:QW_geometry}
\end{figure}

As illustrated in Figure~\ref{fig:LDOS}, the band profile of the SiGe layer between Si resembles a quantum well for electrons and holes. 
Figure~\ref{fig:QW_geometry} schematically introduces the main parameters used to describe such a quantum well. While the wave function is strictly confined inside the infinitely deep quantum well, the finite quantum well allows for an exponentially decaying tail inside the classically forbidden regions.
For the infinite square well of width $t$, the quantized energies $E_n$ are well known to be calculated as
\begin{equation}
	E_n = \frac{\hbar^2}{2 m} { \left( \frac{n \pi}{t} \right)}^2 \quad , \quad n \in \mathbb{N} \quad ,
	\label{eq:quantumwell_infinite}
\end{equation}
where $m$ is the mass of the particle trapped inside the well.
In contrast, the finite square well cannot be solved analytically. However, an implicit solution

\begin{equation}
	\text{tan}\left( \sqrt{\frac{ m E }{ 2 } }\frac{t}{\hbar} \right) = 
	\sqrt{ \frac{ m' } {m} \frac{ (V_0-E) }{ E} }  \quad
	\label{eq:quantumwell_finite}
\end{equation}
can be derived from conditions for continuity and differentiability of the wave function at the interfaces between the three regions. $V_0$ is the depth of the potential, $m$ is the mass inside the well (region II in Figure~\ref{fig:QW_geometry}) and $m'$ is the mass outside the well (regions I and III). This distinction is useful when describing an effective mass in different local material compositions. In the present case of a confined SiGe layer, it is the $z$-component of the effective mass tensor. $V_0$ is given by the band offset between bulk SiGe and Si.

In order to describe the band gap of isolated thin layers, Equation \eqref{eq:quantumwell_infinite} can be generalized into the parameterized form
\begin{equation}
	 E_{\text{gap}} = E_{\text{gap}}^{\text{bulk}} + \frac{\beta}{t^{\alpha}} \quad.
	\label{eq:empirical}
\end{equation}
It provides an excellent fit to simulations of thin layers in vacuum \cite{Dutta_2016,Joseph_2023,Zhang_2017},
which acts as a nearly infinite barrier.

\begin{figure}[hbt]
	\centering
	\includegraphics{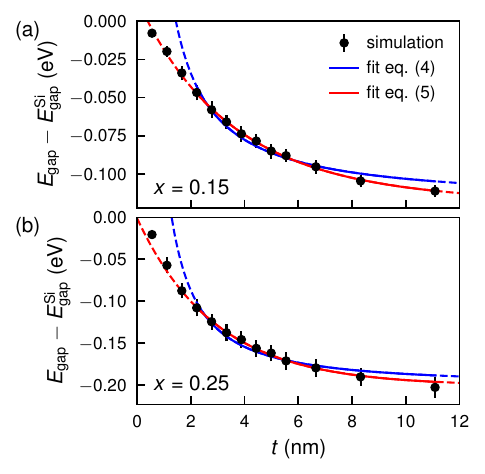}
	\caption{Calculated band gaps of SiGe layers with 15\% and 25\% Ge concentration and standard deviation of 36 atomic structures each. Results of EHT are fitted for $t$~$>$~2~nm with Equations \eqref{eq:empirical} and \eqref{eq:empirical_shift}.}
	\label{fig:bandgap_fitting}
\end{figure}
However, it is not sufficient for describing a confined structure in a surrounding medium, which more closely resembles the finite well. Figure~\ref{fig:bandgap_fitting} illustrates this for two different Ge concentrations: Equation \eqref{eq:empirical} does not fit the EHT results.

Taking the extension of the wave function into account, one can introduce a parameterized shift $\gamma$ into the equation as
\begin{equation}
	 E_{\text{gap}} = E_{\text{gap}}^{\text{bulk}} + \frac{\beta}{ {(t+\gamma)}^{\alpha} } \quad.
	\label{eq:empirical_shift}
\end{equation}

This adjusted formula \eqref{eq:empirical_shift} provides a much better agreement between the fit and the simulation of the atomic structures, as shown by the red curve in Figure~\ref{fig:bandgap_fitting}. 
The concept of introducing an effective layer width in the quantum well model has been previously suggested \cite{Garrett79,Barker91,Rokhsar_1996}.
While the wave function penetration depth in general depends on the energy, a constant value can provide a good approximation. We showcase this fact in greater detail in appendix~\ref{sec:app_approximations}.

\subsection{Describing local stoichiometry fluctuations using the quantum well model}
\label{sec:results_quantumwell}
\begin{figure*}[hbt]
	\centering
	\includegraphics{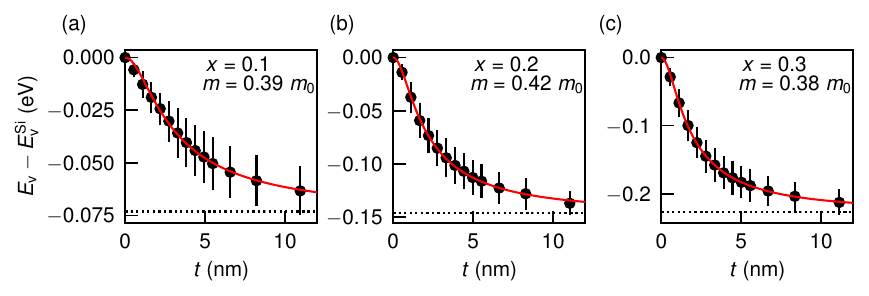}
	\caption{ Valence band edge for thin SiGe layers with three different alloy compositions. The standard deviation of 36 atomic structures is included. The red curve is the finite quantum well model of the same mean depth with fitted mass $m$; the dotted line reflects the band edge position of the bulk alloy as the limit for $t\rightarrow\infty$.}  
	\label{fig:atomistic_and_QW_relativetoSi}
\end{figure*}

Instead of fitting an approximate formula, one can also directly compare EHT results to the numerical solution of the finite potential well.
Using the mass $m$ as a free parameter, we fit the numerical solution of Equation~\eqref{eq:quantumwell_finite} to the EHT calculations. Figure~\ref{fig:atomistic_and_QW_relativetoSi} shows the results. The finite quantum well with a mass of $m\approx(0.40\pm0.02) m_0$ perfectly describes the mean valence band edge of the 36 studied atomic structures.

Figure~\ref{fig:atomistic_and_QW_relativetoSi} also includes the standard deviation of the 36 atomic structures. Given the good agreement of the mean values, one can ask whether the model of the finite well can also be applied to describe the effect of local fluctuations of the atomic structure on the band gap. Two main factors contribute to the fluctuation of band energies between the structures: the different Ge content of each simulation cell and the inherent band gap variation of the alloy.

The first effect originates from the random distribution of Ge atoms in the supercells (see Figure~\ref{fig:supercell}). The Ge content  of the supercell and all sub-cells follow a binomial distribution.
The number of Ge atoms, $N_{\text{Ge}}(t)$, in a layer of $N(t)$ atoms has a variance of 
\begin{equation}
	\sigma^2_{N_{\text{Ge}}(t)} =   (1-x) x N(t) \quad ,
	\label{eq:sigma_N}
\end{equation}
The standard deviation of the Ge content $x=N_{\text{Ge}} / N$ is therefore
\begin{equation}
	 \sigma_x(t) = \sqrt{ \frac{(1-x)x}{N(t)}} \quad .
	 \label{eq:sigma_x}
\end{equation}
This distribution of Ge-content then translates to a distribution of band edges in the ensemble of cells via Equations \eqref{eq:quadratic_all}. 

The second effect is the inherent band gap variation of the alloy. Two atomic structures of the same alloy composition can result in different (local) band gaps. This has been previously studied in SiGe alloys \cite{Dick2025,Fuchs_2025} and other semiconductor alloys as well \cite{Maeder_1995}.

The sum of both effects does not give the observed standard deviation of the confinement energy $E$, but instead defines the variation of the quantum well depth, $V_0$, in different samples. 

We treat both effects to be stochastically independent and simulate an ensemble of 100000 quantum wells with individual depths
\begin{equation}
	V^{i}_0 = E_{\text{v}}(x^i) + \delta^i \quad,
	\label{eq:V0_distribution}
\end{equation}
where the $E_{\text{v}}(x^i)$ is the valence band edge of bulk SiGe as described in Equation \eqref{eq:quadratic_VB}, $x^i$ is given by a continuous uniform distribution of the same standard deviation as  $\sigma_{\text{x}}(t)$ in Equation \eqref{eq:sigma_x}, and $\delta^i$ describes the inherent band gap variation.  $\delta^i$ is uniformly distributed around a mean value of 0, with a standard deviation obtained from previous simulations of bulk SiGe alloys \cite{Dick2025}. We use the standard deviation of the band gap as an approximation for the valence band, as the conduction band offset has little impact on the band gap (see Figure~\ref{fig:LDOS}).

\begin{figure}[hbt]
	\centering
	\includegraphics{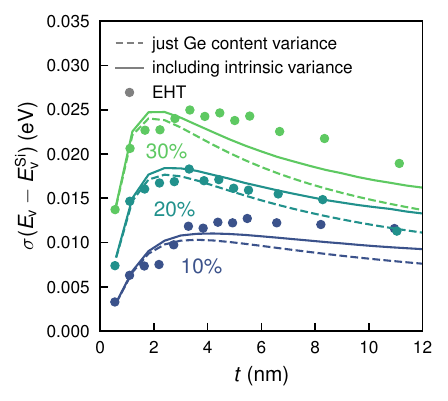}
	\caption{Standard deviation of the band edge for thin SiGe layers with 10\%, 20\%  and 30\% Ge content. Results from EHT are compared to the finite quantum well model. It includes two contributions to the standard deviation, as discussed in the text.}  
	\label{fig:standarddeviation}
\end{figure}
Figure~\ref{fig:standarddeviation} compares the standard deviation of the EHT results and the approximations using the finite quantum well models.

The shape of the curves can be easily understood as the interplay between the distribution of $V_0$ and the fact that the energies $E$ are given relative to the quantum well edge. For small $t\rightarrow0$, the ground state energy is close to the top of the quantum well, i.e.~ $(E^i-V_0^i)\rightarrow0$. The energy difference between quantum wells of different $V_0^i$ is therefore small. It increases with larger $t$ and the distribution of $V_0$ becomes the defining factor. As the layer thickness increases, so does the number of atoms, and consequently the standard deviation decreases (see Equation~\eqref{eq:sigma_x}). This gives rise to a maximum at a certain layer thickness.

The quantum well model captures this effect well. Including the intrinsic band gap variation of the alloy increases the standard deviation and brings the model closer to the EHT results. We estimate the error of the EHT standard deviation to be $\sim$5~meV (see appendix \ref{sec:error_estimation}).
Taking into account that the EHT results are obtained from an ensemble of only 36 atomic structures, as well as the simple model for our distribution of $V_0$ in the quantum wells, they are in excellent agreement.
 
With EHT results giving validity, the quantum well model then allows comparing the individual contributions to the standard deviation of the energies. While the inherent band gap variation is the dominant contribution in bulk-like thick layers, the local variation in Ge content cannot be neglected in the few-nm range.

\FloatBarrier
\section{Conclusions}

We performed EHT calculations for SiGe layers of varying Ge content and layer thicknesses. The  band gaps and band edges of 36 different atomic configurations were calculated to simulate the effect of local stoichiometry fluctuations in real devices. Our results of the ensemble mean values adjust the band gap of bulk SiGe alloys for layer thicknesses in the nm-range. 

The model of the infinite quantum well is not suited to describe quantum confinement in thin SiGe layers when surrounded by another medium with similar band energy levels. Instead, our results are matched by the finite quantum well model, which includes the properties of the surrounding medium into account. By taking into account that the crystal wave function extends beyond the physical size of the well, this introduces the concept of an effective layer thickness, which depends on the band alignment and the effective mass of both interfacing materials. 

The model of the finite quantum well can also be applied to describe the effects of local fluctuations in the atomic structure on the band gap of thin SiGe layers. The band gap variation in few-nm thin layer is larger than in the bulk material and requires consideration. 

Our approach can easily be extended to study quantum confinement in other random alloys, where mean field approaches fail to describe local stoichiometry fluctuations.

\begin{acknowledgments}
	
	The authors thank Fabian Teichert, Angela Thränhardt, Michael Schröter and Prateek Kumar for fruitful discussions.
	This work was funded by the German Research Foundation (Deutsche Forschungsgemeinschaft, DFG) as part of project 466103046 and via project 270/290-1 FUGB, which provided the required computational resources.
	Sibylle Gemming gracefully acknowledges funding by the DFG via project T1 of the Research Unit FOR 5242 (\#449119662).
\end{acknowledgments}

\section*{Data availability}
	The data that support the findings of this article are openly available \cite{Zenodo_thin_layers}.

\appendix
\FloatBarrier
\section{Impact of structural relaxation}

\begin{figure}[hbt]
	\centering
	\includegraphics[width=0.4\textwidth]{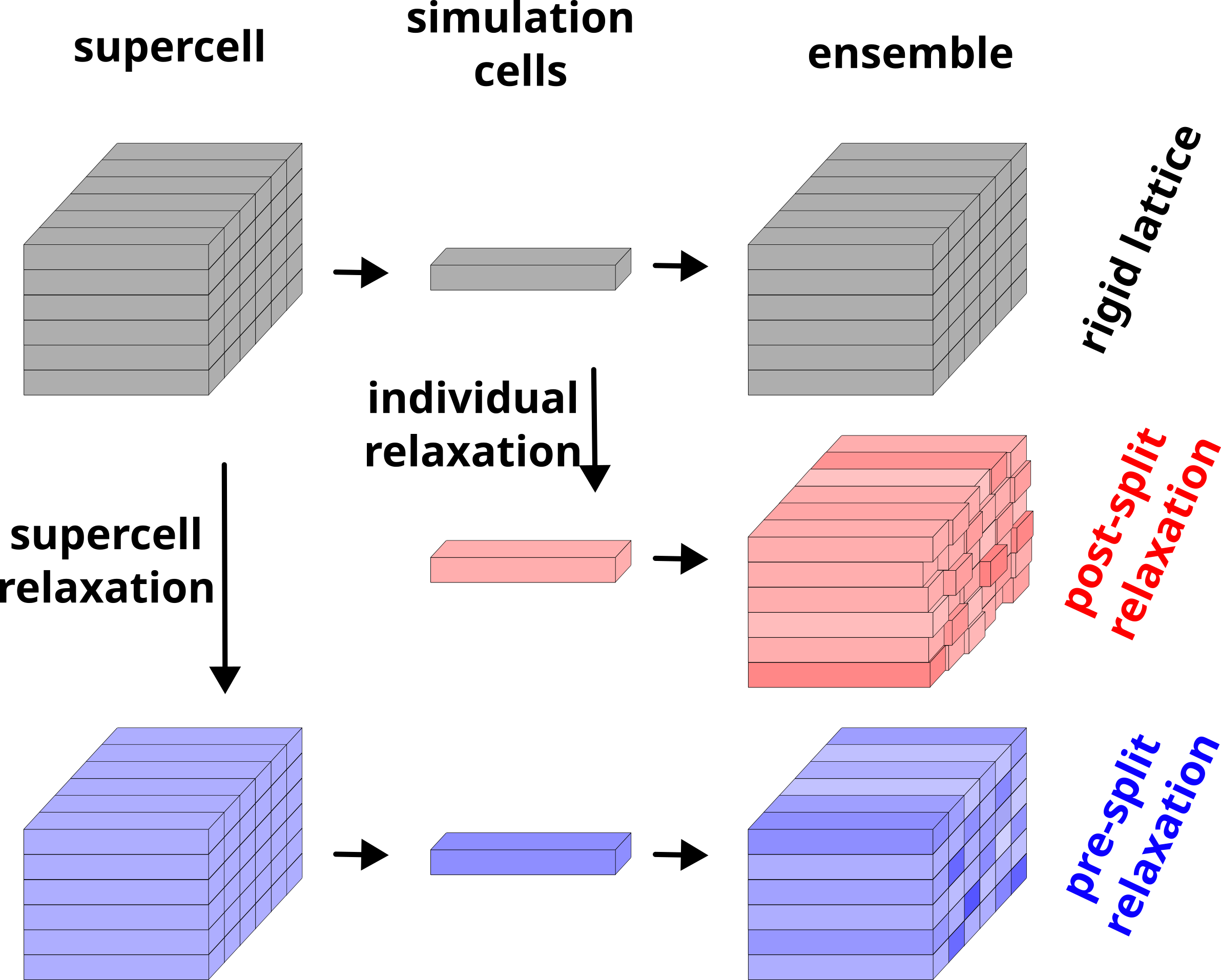}
	\caption{ Three approaches to structural relaxation, as described in the text. Without a relaxation step (grey), the lattice remains unchanged. Individual relaxation of the simulation cells leads to an ensemble with different out-of-plane lattice constants (red). This can be avoided by relaxing the supercell (blue).}
		\label{fig:relaxation}
\end{figure}

In this section we discuss three possible approaches for obtaining the atomic structure for the EHT calculations: rigid lattice, post-split relaxation and pre-split relaxation. Figure~\ref{fig:relaxation} illustrates them.

The most simple option is the \textit{rigid lattice} approach. Knowing the lattice constants of Si and biaxially strained SiGe, one can easily obtain the atom coordinates of an ideal face-centered cubic lattice for any given Ge content. By omitting structural relaxation, this approach is the fastest and defines a common layer thickness for all cells, which makes it easier to compare quantum confinement. However, it fails to describe the effects of local fluctuations of the Ge content on the atomic structure.

By performing \textit{post-split relaxation}, one can include this effect. However, it can lead to different layer thicknesses in each cell, as the individual alloy compositions will not be the same. One can solve this by fixing the cell size in the $r_z$-direction, but this can also lead to unrealistic straining and requires assumptions about the behavior at the Si/SiGe interface.

\textit{Pre-split relaxation} does not rely on such assumptions. By relaxing the supercell of >20,000 atoms, local stress is minimized while the mean Ge content dictates the out-of-plane lattice constant. By splitting up the cells after the relaxation step, the individual simulation cells retain the same thickness. However, the relaxation of the huge supercell is computationally demanding. It becomes feasible through the use of empirical potentials for the relaxation step.

\begin{figure*}[hbt]
	\includegraphics{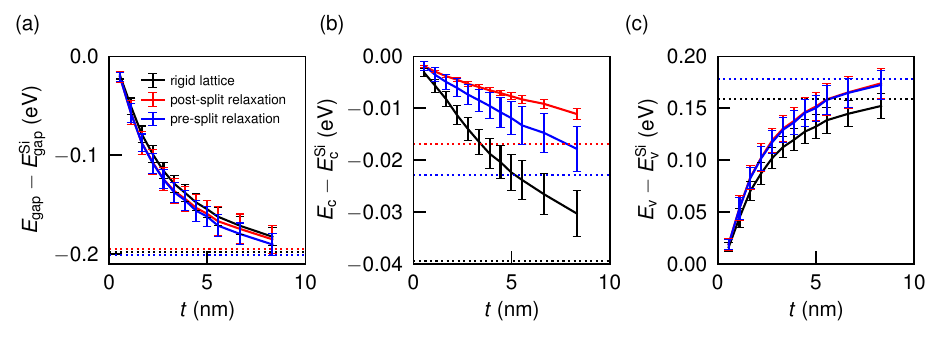}
	\caption{Impact of structural relaxation on the band gap (a), valence band (b), and conduction band (c), using the three approaches discussed in the text. The standard deviation of the 36 different structures is included. The colors match the corresponding steps shown in Figure~\ref{fig:relaxation}. The dotted lines illustrate the corresponding values of the bulk alloy as the limit $t\rightarrow\infty$. Please note the smaller energy range of the conduction band offset.
	}
	\label{fig:bandgap_relaxation}
\end{figure*}

Figure~\ref{fig:bandgap_relaxation} compares the three approaches with regard to the calculated band gap and band offsets in the case of a SiGe alloy with 25\% Ge content. Structural relaxation reduces the band gap slightly and there is a difference between \textit{pre-split} and \textit{post-split relaxation}. Surprisingly, in the case of the bulk alloys (indicated by the dotted line), the ordering is changed. 

Both the conduction band and the valence band (see Figure~\ref{fig:bandgap_relaxation}b and c) show the same orderings for the energies obtained by the three methods. The \textit{rigid lattice} approach yields the lowest energies, overestimating the conduction band offset and underestimating the valence band offset. For the \textit{post-split relaxation}, it is the opposite. \textit{Pre-split relaxation} lies in-between. This is to be expected as it shares the property of a common out-of-plane lattice constant with the former and the local relaxation with the latter.

In the present case, the difference between \textit{pre-split} and \textit{post-split relaxation} is in the order of a few meV. \textit{Post-split relaxation} may therefore be a suitable approach for computationally demanding methods, such as DFT.

Note, however, the strong impact of structural relaxation for the conduction band offset. While the conduction band offset between Si and SiGe is small and structural relaxation therefore has little impact on the absolute value, the relative difference between the three approaches is high.
In cases with different band alignment, e.g.~for strongly n-doped SiGe alloys, \textit{pre-split relaxation} may be required for an accurate description. We noticed this  when using the same band alignment as the original EHT parameterization \cite{Dick2025}, which produced a larger conduction band offset.

\onecolumngrid

\FloatBarrier
\twocolumngrid
\section{Approximations to the finite quantum well model}
\label{sec:app_approximations}

\begin{figure}[hbt]
	\centering
	\includegraphics{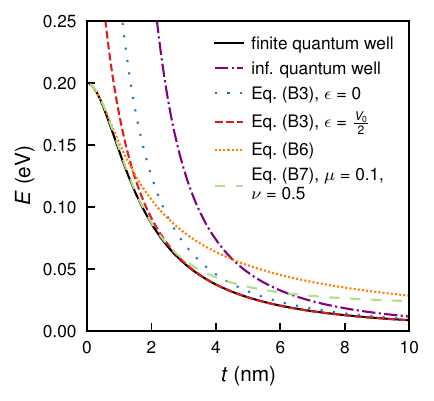}
	\caption{ Comparison between the ground state energies $E$ of the finite and infinite quantum well, given by Equation~\eqref{eq:quantumwell_infinite} and \eqref{eq:quantumwell_finite}, as well as the approximations \eqref{eq:const_dp_approximation}, \eqref{eq:low_t_approximation} and \eqref{eq:empirical_small-t}. We use $m=m'=m_0$ and $V_0$~=~0.2~eV }	
	\label{fig:finite_well_approximations}
\end{figure}

Figure~\ref{fig:finite_well_approximations} compares the ground state energies of the infinite quantum well, the finite quantum well as well as two approximations for the case of small $t$ and large $t$. The following subchapters will introduce and critically evaluate them.

\subsection{Approximations for large $t$}

The main difference between the finite and the infinite quantum well is that the wave function is not strictly confined by the physical dimension in the finite well, but allows for an exponentially decaying tail inside the classically forbidden region (regions I and III in Figure~\ref{fig:QW_geometry}). 

We define the penetration depth $t_{\text{p}}$ by the point where the value of the wave function has decayed to one half of its value at the interface. It is calculated as

\begin{equation}
	t_{\text{p}}(E) = \frac{ ln(2) \hbar }{ \sqrt{ 2m' (V_0-E) }} \quad .
	\label{eq:penetration_depth}
\end{equation}

\begin{figure}[h!bt]
	\centering
	\includegraphics{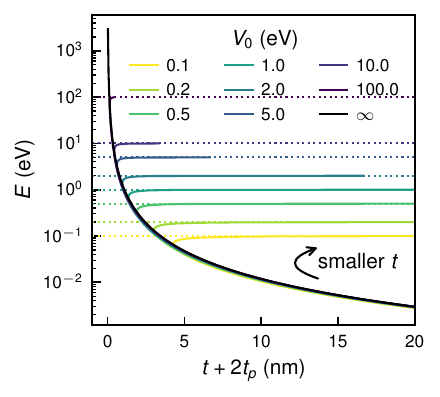}
	\caption{ Comparison between the ground state energies of finite and infinite quantum wells using masses $m=m'=m_0$. The ground state energy $E$ is plotted against the effective well width according to Equation~ \eqref{eq:penetration_depth}. For the infinite well, $t_{\text{p}}$ is assumed to be zero. The dotted line shows the energy at the top of the potential well, $V_0$, as the limit for $t\rightarrow0$.}
	\label{fig:addedpenetrationdepth_half}
\end{figure}

Adding this penetration depth to the physical thickness $t$ gives excellent agreement to the infinite well solution for sufficiently deep wells, as illustrated in Figure~\ref{fig:addedpenetrationdepth_half}. As $t$ is reduced and the ground state energy $E$ approaches the edge of the well, $V_0$, the weakly bound particle has a high penetration depth. This leads to the sharp turn of the curves in Figure~\ref{fig:addedpenetrationdepth_half} and disagreement with the infinite well model, which describes strongly bound particles. In the figure we notice that deeper wells agree with the infinite well model on a wider range of thicknesses, which is to be expected.

\begin{figure}[hbt]
	\centering
	\includegraphics{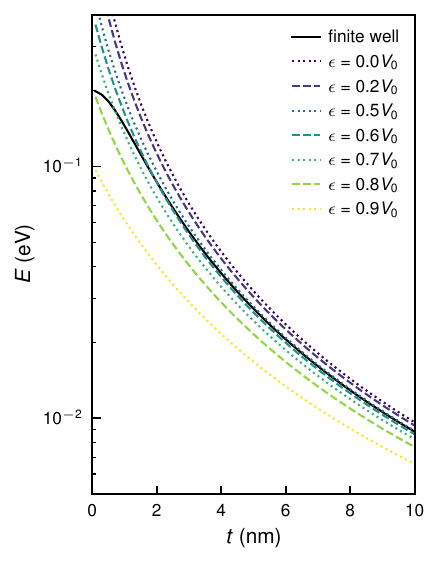}
	\caption{ Comparing the finite quantum well with $m$~=~$m'$~=~$m_0$ and the approximation for large $t$, Equation~\eqref{eq:const_dp_approximation}, for different $\epsilon$. }	
	\label{fig:const_dp_approximation}
\end{figure}

Starting from the solution of the infinite well, Equation \eqref{eq:quantumwell_infinite}, and incorporating Equation \eqref{eq:penetration_depth}, the confinement energy can be approximated as
\begin{equation}
	E \approx \frac{\hbar^2}{2 m} { \left( \frac{n \pi}{t+2t_{\text{p}}(E)} \right)}^2 \quad .
	\label{eq:high_t_approximation}
\end{equation}
Note that this is only an implicit solution for $E(t)$. 
We noticed that while using $t_{\text{p}}(E=0)$ in \eqref{eq:high_t_approximation} gives the correct limit for $t\rightarrow\infty$, a better approximation for moderately thin layers can be found using an adjusted formula
\begin{equation}
	E \approx \frac{\hbar^2}{2 m} { \left( \frac{n \pi}{t+2t_{\text{p}}(\epsilon)} \right)}^2 \quad , \quad \epsilon=\text{const.} \quad ,
	\label{eq:const_dp_approximation}
\end{equation}

where $\epsilon$ can be interpreted as an effective or mean confinement energy across a range of finite layer thicknesses. A value of $\epsilon \approx~50-60\%~V_0$ provides the best agreement with the numerical solution of the finite quantum well, as shown in Figure~\ref{fig:const_dp_approximation}.

\begin{figure}[hbt]
	\centering
	\includegraphics{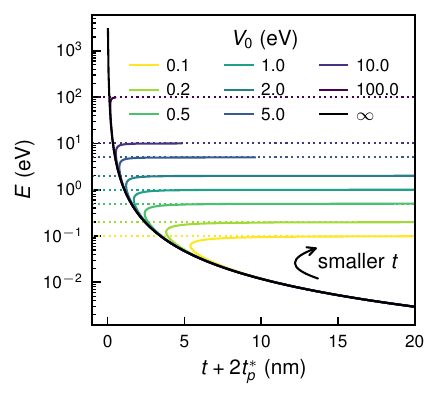}
	\caption{ Comparison between the ground state energies of finite and infinite quantum wells using masses $m=m'=m_0$. The ground state energy $E$ is plotted against the effective well width according to Equation~ \eqref{eq:penetration_depth_e}. For the infinite well, $t_{\text{p}}^*$ is assumed to be zero. The dotted line shows the energy at the top of the potential well, $V_0$, as the limit for $t\rightarrow0$.}
	\label{fig:addedpenetrationdepth_e}
\end{figure}

Other authors \cite{Garrett79, Barker91} have instead defined the penetration depth $t_{\text{p}}^*$ by the point where the value of the wave function has decayed to one over e of its value at the interface. Following this definition, the penetration depth is calculated as
\begin{equation}
	t_{\text{p}}^*(E) = \frac{ \hbar }{ \sqrt{ 2m' (V_0-E) }} \quad .
	\label{eq:penetration_depth_e}
\end{equation}
It differs from Equation~\eqref{eq:penetration_depth} by a factor of ln(2).
Figure~\ref{fig:addedpenetrationdepth_e} illustrates how this alternative definition of the penetration depth provides good agreement with the infinite quantum well. For large $t$, it more closely matches the infinite quantum well than our approach in Figure~\ref{fig:addedpenetrationdepth_half}. However, our definition gives reasonable agreement for a slightly larger range of $t$. 

Barker et al.~\cite{Barker91} approximate a constant penetration depth by using $E=0$ in Equation~\eqref{eq:penetration_depth_e}. This is equivalent to $\epsilon=(1-{\text{ln}(2)}^2) V_0\approx0.51V_0$ in Equation~\eqref{eq:const_dp_approximation}, which lies within the range of good fit as deduced from our model.

\FloatBarrier
\subsection{Approximations for small $t$}

For small $t$, one can use the first order Taylor expansion $\text{tan}(x)=x+O(x^3)$ on the left-hand side of Equation~\eqref{eq:quantumwell_finite}. After squaring both sides and reordering the terms, one obtains the quadratic equations

\begin{equation}
	E^2 + \frac{2\hbar^2}{M t^2} E -  \frac{2 \hbar^2}{M t^2} V_0 =0 
	\label{eq:taylor}
\end{equation}
with solutions
\begin{equation}
	E_{1/2} =  \frac{\hbar^2} {M t^2} \left( \pm \sqrt{ 1 +   \frac{ 2M  V_0  t^2} {\hbar^2} } -1 \right) \quad ,
	\label{eq:low_t_approximation}
\end{equation}

where we use $M=m \frac{m}{m'}$. $M$ can be interpreted as the effective mass in an equivalent well, where $m = m'$. The solution with the negative sign in Equation~\eqref{eq:low_t_approximation} can be ignored.
As shown by the orange curve in Figure~\ref{fig:finite_well_approximations}, the remaining solution provides a good fit to the finite quantum well for thicknesses below 1~nm, but is less useful for larger thicknesses.

Using the Taylor expansion of the square root in Equation~\eqref{eq:low_t_approximation} and further simplifications lead to a generalized form
\begin{equation}
	E = \epsilon + \frac{\mu+ t^2}{\nu + t^2} \quad.
	\label{eq:empirical_small-t}
\end{equation}
The parameters $\mu$ and $\nu$ can then be adjusted to achieve good agreement with the exact solution, as shown in Figure~\ref{fig:finite_well_approximations} by the green curve.

\FloatBarrier
\section{Mean energy of finite quantum wells}
\begin{figure}[hbt]
	\centering
	\includegraphics{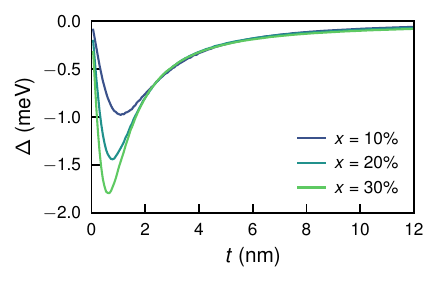}
	\caption{ The relative difference $\Delta$, as defined by Equation \eqref{eq:mean_difference}, as a function of the layer thickness $t$. The role of the Ge content $x$ on the potential well $V_0$ is described in Section ~\ref{sec:results_quantumwell}. }  
	\label{fig:mean_difference}
\end{figure}

We calculate the mean ground state energy for an ensemble of finite quantum wells $i$, with their individual depths $V_0^{i}$ following the distribution described by Equation~\eqref{eq:V0_distribution}. As Figure~\ref{fig:mean_difference} shows, we find that the difference 
\begin{equation}
	\Delta = \langle E(t,V_0^i) - V_0^i\rangle - \left[ E(t, \langle V_0^i\rangle) - \langle V_0^i\rangle  \right] 
	\label{eq:mean_difference}
\end{equation}
between the expected energy of the mean potential $E(t,\langle V_0^i\rangle)$ and the ensemble mean $\langle E(t,V_0^i)\rangle$ is small, but non-zero. Given the corresponding values of $V_0$, the difference is less than 1\%. Depending on the use case, a single simulation of the mean potential well can therefore be sufficient to describe the mean value of the ensemble. The ensemble simulation is primarily required, when further information about the distribution, e.g.,~ the standard deviation, is of interest. We consider this in Section~\ref{sec:results_quantumwell}.

\FloatBarrier
\section{Distribution of the randomly selected Ge content in the simulation cells.}
\label{sec:Ge_content}
\begin{figure}[hbt]
	\centering
	\includegraphics{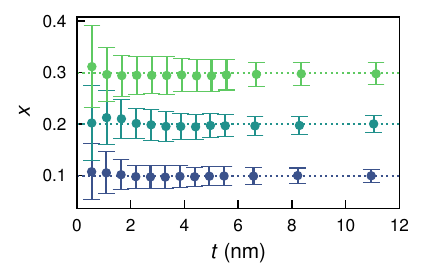}
	\caption{ Mean Ge content $x$ of the ensemble of 36 atomic structures and corresponding standard deviation, shown for three selected alloy compositions.}  
	\label{fig:Ge_content}
\end{figure}

Figure~\ref{fig:Ge_content} shows the distribution of the Ge content in our ensembles of 36 atomic structures. Since the type of each individual atom is randomly chosen, the target alloy composition is only reached in the limit of an infinitely large cell. For finite-sized cells, a small deviation from the expectation value is to be expected. This introduces a small error in our calculations, which we discuss in the next section. It is an artifact of the numerical sample size and can be reduced by averaging a larger number of atomic structures.

\FloatBarrier
\section{Error estimation of EHT results}
\label{sec:error_estimation}

\begin{figure}[hbt]
	\centering
	\includegraphics{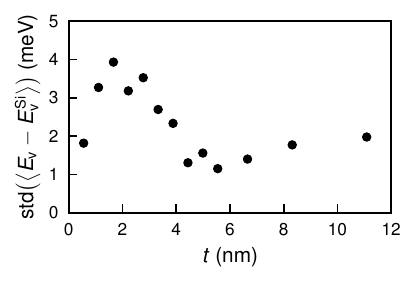}
	\caption{ Variation of the mean valence band position from an ensemble of 36 atomic structures, across 5 ensembles.}  
	\label{fig:error_EHT_mean}
\end{figure}
\begin{figure}[hbt]
	\centering
	\includegraphics{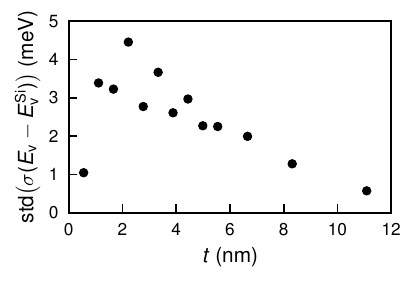}
	\caption{ Variation of the standard deviation of the valence band position from an ensemble of 36 atomic structures, across 5 ensembles.}  
	\label{fig:error_EHT_std}
\end{figure}

Figure~\ref{fig:error_EHT_mean} and \ref{fig:error_EHT_std} show the variation (standard deviation) of the mean valence band energy and its standard deviation in an ensemble of 36 atomic structures, derived from one supercell with roughly 30\% Ge content. For this, we simulate 5 of such ensembles for different layer thicknesses and calculate the sample standard deviation. This allows for a rough estimate of the error in our simulations. The error can be attributed to the random distribution of alloy composition and the fact that the mean Ge content in each ensemble (and its original supercell) is not exactly 30\% (see section \ref{sec:Ge_content}). Both figures show an estimated error of less than 5~meV, for high $t$ it is in the order of $\sim$2~meV.
The shape in both figures resembles that of the standard deviation inside one single ensemble, as show in Figure~\ref{fig:standarddeviation}. This does not imply correlation, but is due to the fact that both, the standard deviation within one ensemble of 36 atomic structures and the standard deviation across multiple of such ensembles, are related to the randomly distributed Ge content. The interplay between the Ge content variation and the depth of the potential well defines the shape of the curve, as discussed in Section~\ref{sec:results_quantumwell}.

\FloatBarrier

\bibliography{USHBER_all}

@Article{Cerda2000,
  author    = {Cerdá, J. and Soria, F.},
  title     = {Accurate and transferable extended Hückel-type tight-binding parameters},
  journal   = {Physical Review B},
  year      = {2000},
  volume    = {61},
  number    = {12},
  pages     = {7965--7971},
  month     = mar,
  issn      = {1095-3795},
  doi       = {10.1103/physrevb.61.7965},
  publisher = {American Physical Society (APS)},
}

@Article{Hoffmann1963,
  author    = {Hoffmann, Roald},
  title     = {An Extended Hückel Theory. I. Hydrocarbons},
  journal   = {The Journal of Chemical Physics},
  year      = {1963},
  volume    = {39},
  number    = {6},
  pages     = {1397--1412},
  month     = sep,
  issn      = {1089-7690},
  doi       = {10.1063/1.1734456},
  publisher = {AIP Publishing},
}

@Article{Braunstein1958,
  author    = {Braunstein, Rubin and Moore, Arnold R. and Herman, Frank},
  title     = {Intrinsic Optical Absorption in Germanium-Silicon Alloys},
  journal   = {Physical Review},
  year      = {1958},
  volume    = {109},
  number    = {3},
  pages     = {695--710},
  month     = feb,
  issn      = {0031-899X},
  doi       = {10.1103/physrev.109.695},
  publisher = {American Physical Society (APS)},
}

@Article{King1989,
  author    = {King, C.A. and Hoyt, J.L. and Gibbons, J.F.},
  title     = {Bandgap and transport properties of Si1-xGex by analysis of nearly ideal Si/Si1-xGex/Si heterojunction bipolar transistors},
  journal   = {IEEE Transactions on Electron Devices},
  year      = {1989},
  volume    = {36},
  number    = {10},
  pages     = {2093--2104},
  issn      = {0018-9383},
  doi       = {10.1109/16.40925},
  publisher = {Institute of Electrical and Electronics Engineers (IEEE)},
}

@Article{Laradji1995,
  author    = {Laradji, Mohamed and Landau, D. P. and Dünweg, B.},
  title     = {Structural properties of Si 1-x Ge x alloys: A Monte Carlo simulation with the Stillinger-Weber potential},
  journal   = {Physical Review B},
  year      = {1995},
  volume    = {51},
  number    = {8},
  pages     = {4894--4902},
  month     = feb,
  issn      = {1095-3795},
  doi       = {10.1103/physrevb.51.4894},
  publisher = {American Physical Society (APS)},
}

@Article{Stillinger1985,
  author    = {Stillinger, Frank H. and Weber, Thomas A.},
  title     = {Computer simulation of local order in condensed phases of silicon},
  journal   = {Physical Review B},
  year      = {1985},
  volume    = {31},
  number    = {8},
  pages     = {5262--5271},
  month     = apr,
  issn      = {0163-1829},
  doi       = {10.1103/physrevb.31.5262},
  publisher = {American Physical Society (APS)},
}

@Article{Lang1985,
  author    = {Lang, D. V. and People, R. and Bean, J. C. and Sergent, A. M.},
  title     = {Measurement of the band gap of Ge x Si 1-x/Si strained-layer heterostructures},
  journal   = {Applied Physics Letters},
  year      = {1985},
  volume    = {47},
  number    = {12},
  pages     = {1333--1335},
  month     = dec,
  issn      = {1077-3118},
  doi       = {10.1063/1.96271},
  publisher = {AIP Publishing},
}

@Article{Smidstrup2019,
  author    = {Smidstrup, Søren and Markussen, Troels and Vancraeyveld, Pieter and Wellendorff, Jess and Schneider, Julian and Gunst, Tue and Verstichel, Brecht and Stradi, Daniele and Khomyakov, Petr A and Vej-Hansen, Ulrik G and Lee, Maeng-Eun and Chill, Samuel T and Rasmussen, Filip and Penazzi, Gabriele and Corsetti, Fabiano and Ojanperä, Ari and Jensen, Kristian and Palsgaard, Mattias L N and Martinez, Umberto and Blom, Anders and Brandbyge, Mads and Stokbro, Kurt},
  journal   = {Journal of Physics: Condensed Matter},
  title     = {QuantumATK: an integrated platform of electronic and atomic-scale modelling tools},
  year      = {2019},
  issn      = {1361-648X},
  month     = oct,
  number    = {1},
  pages     = {015901},
  volume    = {32},
  doi       = {10.1088/1361-648x/ab4007},
  publisher = {IOP Publishing},
}

@Article{Fuchs_2025,
  author    = {Fuchs, Florian and Roscher, Willi and Dick, Daniel and Irmscher, Christian F. and Schuster, Jörg and Gemming, Sibylle},
  title     = {Statistical Studies on Random Configurations of Silicon Germanium Carbon Alloys Using Density Functional Theory},
  journal   = {The Journal of Physical Chemistry C},
  year      = {2025},
  volume    = {129},
  number    = {2},
  pages     = {1546--1552},
  month     = jan,
  issn      = {1932-7455},
  doi       = {10.1021/acs.jpcc.4c08035},
  publisher = {American Chemical Society (ACS)},
}

@Article{Weber_1989,
  author    = {Weber, J. and Alonso, M. I.},
  title     = {Near-band-gap photoluminescence of Si-Ge alloys},
  journal   = {Physical Review B},
  year      = {1989},
  volume    = {40},
  number    = {8},
  pages     = {5683--5693},
  month     = sep,
  issn      = {0163-1829},
  doi       = {10.1103/physrevb.40.5683},
  publisher = {American Physical Society (APS)},
}

@Book{Rinaldi2022,
  title     = {Silicon-Germanium Heterojunction Bipolar Transistors for mm-wave Systems Technology, Modeling and Circuit Applications},
  publisher = {River Publishers},
  year      = {2022},
  author    = {Rinaldi, Niccolò and Schröter, Michael},
  address   = {New York},
  month     = sep,
  isbn      = {9781003339519},
  doi       = {10.1201/9781003339519},
}

@Article{Stokbro_2010,
  author    = {Stokbro, Kurt and Petersen, Dan Erik and Smidstrup, Søren and Blom, Anders and Ipsen, Mads and Kaasbjerg, Kristen},
  title     = {Semiempirical model for nanoscale device simulations},
  journal   = {Physical Review B},
  year      = {2010},
  volume    = {82},
  number    = {7},
  month     = aug,
  issn      = {1550-235X},
  doi       = {10.1103/physrevb.82.075420},
  publisher = {American Physical Society (APS)},
}

@Article{Chu_2023,
  author    = {Chu, Chun-Lin and Hsu, Shu-Han and Chang, Wei-Yuan and Luo, Guang-Li and Chen, Szu-Hung},
  title     = {Stacked SiGe nanosheets p-FET for Sub-3 nm logic applications},
  journal   = {Scientific Reports},
  year      = {2023},
  volume    = {13},
  number    = {1},
  pages     = {9433},
  month     = jun,
  issn      = {2045-2322},
  doi       = {10.1038/s41598-023-36614-2},
  publisher = {Springer Science and Business Media LLC},
}

@Article{Fuchsberger_2023,
  author    = {Fuchsberger, Andreas and Wind, Lukas and Sistani, Masiar and Behrle, Raphael and Nazzari, Daniele and Aberl, Johannes and Prado Navarrete, Enrique and Vukŭsić, Lada and Brehm, Moritz and Schweizer, Peter and Vogl, Lilian and Maeder, Xavier and Weber, Walter M.},
  title     = {Reconfigurable Field‐Effect Transistor Technology via Heterogeneous Integration of SiGe with Crystalline Al Contacts},
  journal   = {Advanced Electronic Materials},
  year      = {2023},
  volume    = {9},
  number    = {6},
  pages     = {2201259},
  month     = apr,
  issn      = {2199-160X},
  doi       = {10.1002/aelm.202201259},
  publisher = {Wiley},
}

@InProceedings{Phillips_2021,
  author    = {Phillips, S. and Preisler, E. and Zheng, J. and Chaudhry, S. and Racanelli, M. and Muller, M. and Schroter, M. and McArthur, W. and Howard, D.},
  title     = {Advances in foundry SiGe HBT BiCMOS processes through modeling and device scaling for ultra-high speed applications},
  booktitle = {2021 IEEE BiCMOS and Compound Semiconductor Integrated Circuits and Technology Symposium (BCICTS)},
  year      = {2021},
  pages     = {1--5},
  month     = dec,
  publisher = {IEEE},
  doi       = {10.1109/bcicts50416.2021.9682485},
}

@InBook{Takagi_2011,
  pages     = {499--527},
  title     = {Silicon–germanium (SiGe)-based field effect transistors (FET) and complementary metal oxide semiconductor (CMOS) technologies},
  publisher = {Elsevier},
  year      = {2011},
  author    = {Takagi, S.},
  isbn      = {9781845696894},
  booktitle = {Silicon–Germanium (SiGe) Nanostructures},
  doi       = {10.1533/9780857091420.4.499},
}

@InBook{Xia_2011,
  pages     = {433--455},
  title     = {Microcavities and quantum cascade laser structures based on silicon–germanium (SiGe) nanostructures},
  publisher = {Elsevier},
  year      = {2011},
  author    = {Xia, J. and Shiraki, Y. and Yu, J.},
  isbn      = {9781845696894},
  booktitle = {Silicon–Germanium (SiGe) Nanostructures},
  doi       = {10.1533/9780857091420.3.433},
}

@Article{Joseph_2023,
  author    = {Joseph, Thomas and Fuchs, Florian and Schuster, Jörg},
  title     = {Electronic structure simulation of thin silicon layers: Impact of orientation, confinement, and strain},
  journal   = {Physica E: Low-dimensional Systems and Nanostructures},
  year      = {2023},
  volume    = {146},
  pages     = {115522},
  month     = jan,
  issn      = {1386-9477},
  doi       = {10.1016/j.physe.2022.115522},
  publisher = {Elsevier BV},
}

@Article{Dutta_2016,
  author    = {Dutta, Tapas and Kumar, Sanjay and Rastogi, Priyank and Agarwal, Amit and Chauhan, Yogesh Singh},
  title     = {Impact of Channel Thickness Variation on Bandstructure and Source-to-Drain Tunneling in Ultra-Thin Body III-V MOSFETs},
  journal   = {IEEE Journal of the Electron Devices Society},
  year      = {2016},
  volume    = {4},
  number    = {2},
  pages     = {66--71},
  month     = mar,
  issn      = {2168-6734},
  doi       = {10.1109/jeds.2016.2522981},
  publisher = {Institute of Electrical and Electronics Engineers (IEEE)},
}

@Article{Khatami_2009,
  author    = {Khatami, Yasin and Banerjee, Kaustav},
  title     = {Steep Subthreshold Slope n- and p-Type Tunnel-FET Devices for Low-Power and Energy-Efficient Digital Circuits},
  journal   = {IEEE Transactions on Electron Devices},
  year      = {2009},
  volume    = {56},
  number    = {11},
  pages     = {2752--2761},
  month     = nov,
  issn      = {1557-9646},
  doi       = {10.1109/ted.2009.2030831},
  publisher = {Institute of Electrical and Electronics Engineers (IEEE)},
}

@Article{Zhang_2017,
  author    = {Zhang, Mi and Xu, Maji and Li, Mingkai and Zhang, Qingfeng and Lu, Yinmei and Chen, Jingwen and Li, Ming and Dai, Jiangnan and Chen, Changqing and He, Yunbin},
  title     = {SnO2 epitaxial films with varying thickness on c-sapphire: Structure evolution and optical band gap modulation},
  journal   = {Applied Surface Science},
  year      = {2017},
  volume    = {423},
  pages     = {611--618},
  month     = nov,
  issn      = {0169-4332},
  doi       = {10.1016/j.apsusc.2017.06.250},
  publisher = {Elsevier BV},
}

@Article{Schaeffler_1997,
  author    = {Schäffler, Friedrich},
  title     = {High-mobility Si and Ge structures},
  journal   = {Semiconductor Science and Technology},
  year      = {1997},
  volume    = {12},
  number    = {12},
  pages     = {1515--1549},
  month     = dec,
  issn      = {1361-6641},
  doi       = {10.1088/0268-1242/12/12/001},
  publisher = {IOP Publishing},
}

@InBook{Jungemann2003_chapter6,
  chapter   = {6.2},
  pages     = {103},
  title     = {Hierarchical Device Simulation},
  publisher = {Springer},
  year      = {2003},
  author    = {Jungemann, Christoph},
  editor    = {Bernd Meinerzhagen},
  series    = {Computational Microelectronics},
  address   = {Vienna},
  isbn      = {9783709160862},
  pagetotal = {254147},
  ppn_gvk   = {749266856},
  subtitle  = {The Monte-Carlo Perspective},
}

@Article{Shiraki_2005,
  author    = {Shiraki, Yasuhiro and Sakai, Akira},
  title     = {Fabrication technology of SiGe hetero-structures and their properties},
  journal   = {Surface Science Reports},
  year      = {2005},
  volume    = {59},
  number    = {7–8},
  pages     = {153--207},
  month     = nov,
  issn      = {0167-5729},
  doi       = {10.1016/j.surfrep.2005.08.001},
  publisher = {Elsevier BV},
}

@Article{Thewalt_1997,
  author    = {Thewalt, M. L. W. and Harrison, D. A. and Reinhart, C. F. and Wolk, J. A. and Lafontaine, H.},
  title     = {Type II Band Alignment in Si 1-x Ge x / Si (001) Quantum Wells: The Ubiquitous Type I Luminescence Results from Band Bending},
  journal   = {Physical Review Letters},
  year      = {1997},
  volume    = {79},
  number    = {2},
  pages     = {269--272},
  month     = jul,
  issn      = {1079-7114},
  doi       = {10.1103/physrevlett.79.269},
  publisher = {American Physical Society (APS)},
}

@Article{Houghton_1995,
  author    = {Houghton, D. C. and Aers, G. C. and Yang, S.-R. Eric and Wang, E. and Rowell, N. L.},
  title     = {Type I Band Alignment in Si 1-x Ge x /Si(001) Quantum Wells: Photoluminescence under Applied [110] and [100] Uniaxial Stress},
  journal   = {Physical Review Letters},
  year      = {1995},
  volume    = {75},
  number    = {5},
  pages     = {866--869},
  month     = jul,
  issn      = {1079-7114},
  doi       = {10.1103/physrevlett.75.866},
  publisher = {American Physical Society (APS)},
}

@Article{Virgilio_2006,
  author    = {Virgilio, Michele and Grosso, Giuseppe},
  title     = {Type-I alignment and direct fundamental gap in SiGe based heterostructures},
  journal   = {Journal of Physics: Condensed Matter},
  year      = {2006},
  volume    = {18},
  number    = {3},
  pages     = {1021--1031},
  month     = jan,
  issn      = {1361-648X},
  doi       = {10.1088/0953-8984/18/3/018},
  publisher = {IOP Publishing},
}

@Article{Schroter_2011,
  author    = {Schröter, Michael and Wedel, Gerald and Heinemann, Bernd and Jungemann, Christoph and Krause, Julia and Chevalier, Pascal and Chantre, Alain},
  title     = {Physical and Electrical Performance Limits of High-Speed SiGeC HBTs—Part I: Vertical Scaling},
  journal   = {IEEE Transactions on Electron Devices},
  year      = {2011},
  volume    = {58},
  number    = {11},
  pages     = {3687--3696},
  month     = nov,
  issn      = {1557-9646},
  doi       = {10.1109/ted.2011.2163722},
  publisher = {Institute of Electrical and Electronics Engineers (IEEE)},
}

@Article{Fuchs_2019,
  author    = {Fuchs, Florian and Gemming, Sibylle and Schuster, Jörg},
  title     = {Radially resolved electronic structure and charge carrier transport in silicon nanowires},
  journal   = {Physica E: Low-dimensional Systems and Nanostructures},
  year      = {2019},
  volume    = {108},
  pages     = {181--186},
  month     = apr,
  issn      = {1386-9477},
  doi       = {10.1016/j.physe.2018.12.002},
  publisher = {Elsevier BV},
}

@Article{Tan_2015,
  author    = {Tan, Yaohua P. and Povolotskyi, Michael and Kubis, Tillmann and Boykin, Timothy B. and Klimeck, Gerhard},
  title     = {Tight-binding analysis of Si and GaAs ultrathin bodies with subatomic wave-function resolution},
  journal   = {Physical Review B},
  year      = {2015},
  volume    = {92},
  number    = {8},
  pages     = {085301},
  month     = aug,
  issn      = {1550-235X},
  doi       = {10.1103/physrevb.92.085301},
  publisher = {American Physical Society (APS)},
}

@Article{Kharche_2008,
  author    = {Kharche, Neerav and Luisier, Mathieu and Boykin, Timothy B. and Klimeck, Gerhard},
  title     = {Electronic structure and transmission characteristics of SiGe nanowires},
  journal   = {Journal of Computational Electronics},
  year      = {2008},
  volume    = {7},
  number    = {3},
  pages     = {350--354},
  month     = jan,
  issn      = {1572-8137},
  doi       = {10.1007/s10825-008-0191-9},
  publisher = {Springer Science and Business Media LLC},
}

@Article{Iori_2014,
  author    = {Iori, Federico and Ossicini, Stefano and Rurali, Riccardo},
  title     = {Structural and electronic properties of Si1-xGex alloy nanowires},
  journal   = {Journal of Applied Physics},
  year      = {2014},
  volume    = {116},
  number    = {15},
  pages     = {154301},
  month     = oct,
  issn      = {1089-7550},
  doi       = {10.1063/1.4898130},
  publisher = {AIP Publishing},
}

@Article{Martinez_Blanque_2014,
  author    = {Martinez-Blanque, Celso and Ruiz, Francisco G. and Godoy, Andres and Marin, Enrique G. and Donetti, Luca and Gámiz, Francisco},
  title     = {Influence of alloy disorder scattering on the hole mobility of SiGe nanowires},
  journal   = {Journal of Applied Physics},
  year      = {2014},
  volume    = {116},
  number    = {24},
  pages     = {244502},
  month     = dec,
  issn      = {1089-7550},
  doi       = {10.1063/1.4904856},
  publisher = {AIP Publishing},
}

@Article{Garrett79,
  author    = {Garrett, Steven},
  title     = {Bound state energies of a particle in a finite square well: A simple approximation},
  journal   = {American Journal of Physics},
  year      = {1979},
  volume    = {47},
  number    = {2},
  pages     = {195--195},
  month     = feb,
  issn      = {1943-2909},
  doi       = {10.1119/1.11875},
  publisher = {American Association of Physics Teachers (AAPT)},
}

@Article{Barker91,
  author    = {Barker, Barry I. and Rayborn, Grayson H. and Ioup, Juliette W. and Ioup, George E.},
  title     = {Approximating the finite square well with an infinite well: Energies and eigenfunctions},
  journal   = {American Journal of Physics},
  year      = {1991},
  volume    = {59},
  number    = {11},
  pages     = {1038--1042},
  month     = nov,
  issn      = {1943-2909},
  doi       = {10.1119/1.16644},
  publisher = {American Association of Physics Teachers (AAPT)},
}

@Article{Rokhsar_1996,
  author    = {Rokhsar, D. S.},
  title     = {Ehrenfest’s theorem and the particle-in-a-box},
  journal   = {American Journal of Physics},
  year      = {1996},
  volume    = {64},
  number    = {11},
  pages     = {1416--1418},
  month     = nov,
  issn      = {1943-2909},
  doi       = {10.1119/1.18367},
  publisher = {American Association of Physics Teachers (AAPT)},
}

@Article{Maeder_1995,
  author    = {Mäder, Kurt A. and Zunger, Alex},
  title     = {Short- and long-range-order effects on the electronic properties of III-V semiconductor alloys},
  journal   = {Physical Review B},
  year      = {1995},
  volume    = {51},
  number    = {16},
  pages     = {10462--10476},
  month     = apr,
  issn      = {1095-3795},
  doi       = {10.1103/physrevb.51.10462},
  publisher = {American Physical Society (APS)},
}

@Article{Dick2025,
  author    = {Dick, Daniel and Fuchs, Florian and Schuster, Jörg and Gemming, Sibylle},
  title     = {An Extended Hückel Theory Parameter Set for Efficient Electronic Structure Calculations of SiGe Alloys},
  journal   = {physica status solidi (RRL) – Rapid Research Letters},
  year      = {2025},
  volume    = {20},
  number    = {2},
  month     = may,
  issn      = {1862-6270},
  doi       = {10.1002/pssr.202500087},
  publisher = {Wiley},
}

@Article{Tunica_2026,
  author    = {Túnica, Marc and Chiodi, Francesca and Amato, Michele},
  title     = {Structural and thermodynamic stability in hexagonal-diamond Si1-x-yGexByalloys},
  journal   = {Semiconductor Science and Technology},
  year      = {2026},
  volume    = {41},
  number    = {1},
  pages     = {015005},
  month     = jan,
  issn      = {1361-6641},
  doi       = {10.1088/1361-6641/ae30f8},
  publisher = {IOP Publishing},
}

@Misc{Zenodo_thin_layers,
  author       = {Dick, Daniel and Fuchs, Florian and Schuster, Jörg and Gemming, Sibylle},
  title        = {Data related to paper "Quantum confinement in semiconductor random alloys: a case study on Si/SiGe/Si" [Data set]},
  howpublished = {Zenodo, doi: 10.5281/zenodo.18873175},
  month        = {03},
  year         = {2026},
  doi          = {10.5281/zenodo.18873175},
  url          = {https://doi.org/10.5281/zenodo.18873175},
}

\end{document}